\documentclass[11pt]{article}

\let\phi\varphi
\let\epsilon\varepsilon
\let\rho\varrho

\usepackage{color}
\usepackage[utf8]{inputenc}
\usepackage[english]{babel}
\usepackage[T1]{fontenc}
\usepackage[margin=1in]{geometry}
\usepackage[activate=normal]{pdfcprot}

\usepackage{ifpdf}
\ifpdf
  \usepackage[pdftex]{graphicx}
  \graphicspath{{./Fig_pdf/}{./Pdf/}}
  \DeclareGraphicsExtensions{.pdf}
\else
  \usepackage[dvips]{graphicx}
  \graphicspath{{./Fig_eps/}{./Eps/}}
  \DeclareGraphicsExtensions{.ps,.eps}
\fi

\usepackage{amsmath}
\usepackage{amssymb}
\usepackage{amsfonts}
\usepackage{latexsym}
\usepackage{xspace}
\usepackage{scrtime}
\usepackage{mdwlist}
\usepackage{enumerate}

\usepackage[algoruled,longend,vlined]{algorithm2e}

\usepackage[amsmath,thmmarks,noconfig]{ntheorem}
\newtheorem{lemma}{Lemma}
\newtheorem{theorem}[lemma]{Theorem}

\qedsymbol{\quad\ensuremath{\Box}}

\ifpdf
  \usepackage[pdftex]{hyperref}
\else
  \usepackage[dvips]{hyperref}
\fi

\usepackage{listings}
\DeclareMathOperator{\OPT}{OPT}
\DeclareMathOperator{\GBC}{GBC}

\title{Maximum Betweenness Centrality:\\ Approximability and Tractable Cases}

\author{
Martin Fink\\ 
Chair of Computer Science~I\\ 
University of W\"urzburg\\
martin.a.fink@uni-wuerzburg.de
\and 
Joachim Spoerhase\\
Chair of Computer Science~I\\
University of W\"urzburg\\
joachim.spoerhase@uni-wuerzburg.de
}

\newcommand{\gbc}{\operatorname{GBC}}

\begin{document}

\maketitle

\begin{abstract}
  The \textsc{Maximum Betweenness Centrality} problem (MBC) can be
  defined as follows.  Given a graph find a $k$-element
  node set $C$ that maximizes the probability of detecting
	communication between a pair of nodes $s$ and $t$ chosen uniformly at random.
	It is assumed that the communication between $s$ and $t$ is
  realized along a shortest $s$--$t$ path which is, again, selected uniformly
  at random.  The communication is detected if the communication path
  contains a node of $C$.

  Recently, Dolev et al.\ (2009) showed that MBC is NP-hard and gave a
  $(1-1/e)$-approximation using a greedy approach.  We provide a
  reduction of MBC to \textsc{Maximum Coverage} that simplifies the
	analysis of the algorithm of Dolev et al.\ considerably.  Our reduction allows
	us to obtain a new algorithm with the same approximation ratio for a
	(generalized) budgeted version of MBC.  We provide tight examples showing that
	the analyses of both algorithms are best possible.  Moreover, we prove that
	MBC is APX-complete and provide an exact polynomial-time algorithm for MBC on
	tree graphs.
\end{abstract}

\section{Introduction}

A question that frequently arises in the analysis of complex networks
is how \emph{central} or \emph{important} a given node is.  Examples
of such complex networks are communication or logistical networks.
There is a multitude of different measures of centrality known in the
literature.  Many of these measures are based on distances.  Consider,
for example, the measures used for the center or the median location
problem.  We, in contrast, are interested in centrality measures that aim at
monitoring communication or traffic.

We investigate a centrality measure called \emph{shortest path betweenness
centrality}
\cite{freeman:freeman-measures-centrality,brandes:variants-of-betweenness}.
This measure can be motivated by the
following scenario that relies only on very basic assumptions.
Communication occurs between a pair $(s,t)$ of distinct nodes that is
selected uniformly at random among all node pairs.  The communication
is always established along a shortest $s$--$t$ path where each such
path is chosen with equal probability.  The centrality of a node $v$
is defined as the probability of detecting the communication, that is,
the probability that $v$ lies on the communication path.

As a possible application we refer to the task of placing a server in
a computer network so as to maximize the probability of detecting
malicious data packets.  Another example is the deployment of toll
monitoring systems in a road network.

As suggested by the previous application example, a natural extension of the
above scenario is to measure the probability of detecting communication for a
whole set of nodes.  The resulting centrality measure is called \emph{group
betweenness centrality}
\cite{everett-borgatti:centrality-groups,dolev+etal:incremental-deployment-group-betweenness}.

In this paper we investigate the problem of finding a given number $k$
of nodes such that the group betweenness centrality is maximized.  We
call this problem \textsc{Maximum Betweenness Centrality} (MBC).




\paragraph*{Previous Results}

The shortest path betweenness centrality was introduced by Freeman
\cite{freeman:freeman-measures-centrality}.  Brandes
\cite{brandes:faster-algorithm-centrality,brandes:variants-of-betweenness}
and Newman \cite{newman:shorest-path-centrality} independently
developed the same algorithm for computing the shortest path betweenness
centrality of all nodes in $O(nm)$ time.

Group betweenness centrality was introduced by Everett and Borgatti
\cite{everett-borgatti:centrality-groups}.  Puzis et al.\
\cite{puzis+etal:successive-computation-group-betweenness-centrality}
gave an algorithm for computing the group betweenness centrality of a
given node set that runs in $O(n^3)$.

Puzis et al.\ \cite{puzis+etal:prominent-group-complex-networks}
introduced MBC, that is, the problem of finding a $k$-element node set
maximizing the group betweenness centrality.  They showed that the
problem is NP-hard. They also gave a greedy algorithm
\cite{puzis+etal:prominent-group-complex-networks,
puzis+etal:successive-computation-group-betweenness-centrality} and
showed that their algorithm yields an approximation factor of $1 - 1/e$
\cite{dolev+etal:incremental-deployment-group-betweenness}.  We remark
that Puzis et al.\ used the name KPP-Com instead of MBC.


\paragraph*{Our Contribution} We provide a reduction from MBC to the well-known
\textsc{Maximum Coverage} problem which we define in Section
\ref{sec:approximation}. This reduction yields a much simpler proof of the
approximability result of Dolev et al.\
\cite{dolev+etal:incremental-deployment-group-betweenness}.  Our
reduction also allows us to derive a new algorithm for a budgeted
version of the problem, which achieves the same approximation factor.
One remarkable property of our reduction is that it is \emph{not} a
polynomial time reduction.  Rather, the reduction is carried out
implicitly and aims at analyzing the algorithms.

We show that the analyses of these algorithms cannot be improved by
providing tight examples (see Section \ref{sec:tight_examples}).  We also prove
that MBC is APX-complete thereby showing that MBC does not admit a PTAS
(Section \ref{sec:apx-completeness}).

Finally, we develop an exact polynomial-time algorithm for MBC on tree
graphs (see Section \ref{sec:polyn-time-algor}).

\paragraph*{Problem Definition}



The input of MBC is an undirected and connected graph $G=(V,E)$ with
node costs $c\colon V\rightarrow\mathbb{R}^+_0$ and a budget $b$.  Let
$s,t \in V$ be the two communicating nodes. By $\sigma_{s,t}$
we denote the number of shortest paths between $s$ and $t$. For $C
\subseteq V$ let $\sigma_{s,t}(C)$ be the number of shortest $s$--$t$
paths containing at least one node of $C$.  So $C$ detects the
communication of $s$ and $t$ with probability
$\sigma_{s,t}(C)/\sigma_{s,t}$ since we assume that the communication
path is selected uniformly at random among all shortest $s$--$t$
paths.  As the selection of any node pair as the communicating pair
$(s,t)$ is equally likely, the probability that $C$ detects the
communication is proportional to the sum
\[\gbc(C) := \sum_{s,t \in V
  \mid s \neq t} \frac{\sigma_{s,t}(C)}{\sigma_{s,t}}\]
which is called \emph{Group Betweenness Centrality}.  The \textsc{Maximum
Betweenness Centrality} problem consists in finding a set $C \subseteq V$ with
$c(C)\leq b$ such that the group betweenness centrality $\gbc(C)$ is
maximized.

\section{Approximation Algorithms}
\label{sec:approximation}

\paragraph*{The Reduction}
Dolev et al.\
\cite{dolev+etal:incremental-deployment-group-betweenness} prove the
approximation factor of their algorithm by a technique inspired by a
proof of the same factor for the greedy algorithm for the well-known
\textsc{Maximum Coverage} problem \cite{feige:set-cover-threshold}.

In what follows we give a reduction to \textsc{Budgeted Maximum
Coverage} \cite{Khuller+etal:budgeted-max-coverage} which is defined
as follows.  The input is a set $S$ of ground elements with weight
function $w \colon S \to \mathbb{R}_{0}^{+}$, a family $\mathcal{F}$
of subsets of $S$, costs ${c'\colon\mathcal{F}\rightarrow\mathbb{R}^+_0}$
and a budget $b \ge 0$.  The goal is to find a collection $C' \subseteq
\mathcal{F}$ with $c'(C') \le b$ such that the total weight $w(C')$ of
ground elements covered by $C'$ is maximized.

The idea of our reduction is to model every shortest path of the graph
$G$ by a ground element with a corresponding weight. Every node $v$ of
$G$ is modeled by the set of (ground elements corresponding to) shortest paths
that contain $v$.

Let $\left( G = (V,E), c, b \right)$ be an instance of MBC.  Let
$S(G)$ be the set of all shortest $s$--$t$ paths between pairs $s,t$
of distinct nodes.  For a shortest $s$--$t$ path $P$ let $w(P) :=
1/\sigma_{s,t}$ be its weight.

For a node $v$ let $S(v)$ be the set of all shortest paths containing
$v$. Set $c'(S(v)) := c(v)$. Finally let $\mathcal{F}(G) := \{\,
S(v)\mid v\in V\,\}$ be our family of sets.  This completes the
construction of our instance $(S(G), w, \mathcal{F}(G), c',b)$ of
\textsc{Budgeted Maximum Coverage}.

Let $C\subseteq V$ be a set of nodes.  Then $S(C):=\bigcup_{v\in C}S(v)$ denotes
the set of all shortest paths containing at least one node of $C$.  It is not
hard to check that
\begin{displaymath}
	w(S(C)) =  \sum_{{s,t\in V\mid s \neq t}} \frac{1}{\sigma_{s,t}} \cdot
	\sigma_{s,t}(C) = \gbc(C)
\end{displaymath}
holds.  Therefore, the group betweenness centrality of a set of nodes
equals the weight of the corresponding set of shortest paths in the
maximum coverage instance. Of course the feasible solutions of MBC and
the feasible solutions of the reduced instance of \textsc{Maximum
  Coverage} are in 1-1-correspondence and have the same goal function
value. Hence corresponding feasible solutions have also the same
approximation ratio for the respective problem instances.  We will
exploit this fact to turn approximation algorithms for
\textsc{Maximum Coverage} into approximation algorithms for MBC with
the same approximation ratio, respectively.  We note, however, that
the reduction is not polynomial.

\paragraph*{The Unit-Cost Version}
First we consider the unit cost variant of MBC, that is, $c\equiv 1$, which has
been introduced by Dolev et al.\
\cite{puzis+etal:prominent-group-complex-networks}.

Consider an instance of unit-cost MBC.  Then the reduction of the
previous section yields an instance of unit-cost \textsc{Maximum Coverage}.  It
is well-known that a natural greedy approach has an approximation factor of
$1-1/e$ for unit-cost \textsc{Maximum Coverage}
\cite{feige:set-cover-threshold}.  The greedy algorithm works as follows:  Start
with an empty set $C'$ and then iteratively add to $C'$ the set $S'\in\mathcal
F$ that maximizes $w(C'+S')$.

Now let's turn back to MBC.  Of course, we do not obtain an efficient
algorithm if we apply the above greedy algorithm explicitly to the
instance of \textsc{Maximum Coverage} constructed by our reduction
since this instance might be exponentially large.  If we, however,
translate the greedy approach for \textsc{Maximum Coverage} back to
MBC we arrive at the following algorithm: Start with an empty node set
$C$ and then iteratively add to $C$ the node $v$ that maximizes
$\gbc(C+v)$.  Observe that the greedy algorithm for \textsc{Maximum
  Coverage} and the greedy algorithm for MBC produce feasible
solutions that are corresponding to each other according to our
reduction. Hence the latter algorithm has an approximation ratio of
$1-1/e$, too.

An implementation of the greedy approach for MBC outlined before has
been developed by Dolev et al.\
\cite{puzis+etal:prominent-group-complex-networks,puzis+etal:successive-computation-group-betweenness-centrality,dolev+etal:incremental-deployment-group-betweenness}.
The authors, however, carry out the analysis of its approximation performance
from scratch inspired by the analysis of Feige \cite{feige:set-cover-threshold}
for \textsc{Maximum Coverage}.

The crucial point in the implementation of Dolev et al.\
\cite{puzis+etal:prominent-group-complex-networks,puzis+etal:successive-computation-group-betweenness-centrality} is, given a node
set $C$, how to determine a node $v$ maximizing $\gbc(C+v)$.  The main
idea of their algorithm is to maintain a data structure that allows to
obtain the value $\gbc(C + v)$ for any $v\in V$ in $O(1)$ time where
$C$ is the set of nodes that the greedy algorithm has chosen so far. An update
of their data structure takes $O(n^{2})$ time if a node $v$ is added to
$C$.  The total running time of all greedy steps is therefore $O(kn^{2})$. This
running time is dominated by $O(n^{3})$ time needed for a preprocessing step for
the initialization of their data structure.

\paragraph*{The Budgeted Version}
The natural generalization of the greedy approach to \textsc{Budgeted
  Maximum Coverage} would add in each greedy step a set $S'$ that
maximizes the relative gain $(w(C'+S')-w(C'))/c(S')$ among all sets
that respect the budget bound, that is, $c(C'+S')\leq b$.  Here, $C'$
is the collection of sets already selected.

As shown by Khuller et al.\ \cite{Khuller+etal:budgeted-max-coverage}
this simple approach achieves an approximation factor of
$1-1/\sqrt{e}$ ($\approx 0.39$) in the case of arbitrary costs. The authors,
however,
give a modified greedy algorithm with an approximation factor of
$1-1/e$ ($\approx 0.63$). The difference to the naive approach is not to start with an
empty set $C'$ but to try all initializations of $C'$ with at most
three sets of $\mathcal F$ that respect the budget bound $b$.  Each of
these initializations is then augmented to a candidate solution using
the above greedy steps.  The algorithm chooses the best among the
candidate solutions. 

By means of our reduction, we transform this algorithm into an
algorithm for budgeted MBC that has the same approximation ratio (confer
Algorithm \ref{algo_approx_mbc}).  We
start with every set of at most three nodes $C \subseteq V$ not
exceeding the budget and then enlarge this set using greedy steps.
Given such a node set $C$, each greedy step selects the node $v$ that
maximizes the relative gain $(\gbc(C +v) - \gbc(C))/c(v)$ among all
nodes that respect the budget bound, that is, $c(C+v)\leq b$.  Finally
the algorithm chooses the best candidate solution found.  Our
reduction proves that the approximation performance of this algorithm
is again $1-1/e$.

\begin{algorithm}[ht]
	\DontPrintSemicolon
	\SetArgSty{}
	\caption{Greedy-Algorithm for MBC}
	\KwIn{$G=(V,E),c,b$}
	\label{algo_approx_mbc}

	$H := \emptyset$\;

	\ForEach{$C \subseteq V$ with $|C| \le 3$ and $c(C) \le b$}{
		$U := V \setminus C$\;
		\While{$U \neq \emptyset$}{
		$u := \operatorname{arg\,max}_{v \in U} \frac{\gbc(C +v) -
		\gbc(C)}{c(v)}$\;
		\If{$c(C + u) \le b$}{$C := C + u$\;}
		$U := U -u$\;
		}
		\lIf{$\gbc(C) > \gbc(H)$}{$H := C$\;}
	}

	\Return $H$\;
\end{algorithm}

It remains to explain how a greedy step is implemented.  As in the
unit-cost case we can employ the data structure of Dolev et al.\
\cite{puzis+etal:successive-computation-group-betweenness-centrality}
that allows to obtain the value $\gbc(C+v)$ in $O(1)$ time.  Since we
know $\gbc(C)$ from the previous step, we can also compute the relative
gain $(\gbc(C +v) - \gbc(C))/c(v)$ for each node $v\in V$ in constant
time.

As the update time of the data structure is $O(n^{2})$ when the set
$C$ is augmented by a node $v$ we get a running time of $O(n^3)$ for
the augmentation stage for any fixed initialization of $C$.  Since
there are at most $O(n^3)$ initializations and the preprocessing of
the data structure takes $O(n^3)$ time we obtain a total running time
of $O(n^6)$.

The simpler greedy approach (which only tests the initialization $C =
\emptyset$) can of course also be adopted for budgeted MBC.  This
algorithm runs in $O(n^3)$ time and has, as mentioned above, an
approximation factor of $1-1/\sqrt{e}$ (and $1-1/e$ in the case of
unit costs).

\begin{theorem}
	There is an $O(n^{3})$-time factor-$(1-1/\sqrt{e})$ and an $O(n^{6})$-time
	factor-$(1-1/e)$ approximation algorithm for \textsc{Maximum
	Betweenness Centrality}. \qed{}
\end{theorem}

\section{Tight Examples}
\label{sec:tight_examples}

Feige \cite{feige:set-cover-threshold} showed that even the unit-cost
\textsc{Maximum Coverage} problem is not approximable within an approximation
factor better than $1 - 1/e$ thereby showing that the greedy algorithm is
optimal in terms of the approximation ratio.  This lower bound, however, does
not carry over immediately to MBC because we have only a reduction from MBC to
\textsc{Maximum Coverage} and not the other way round.

In what follows we provide a class of tight examples and thus show
that the \emph{analyses} of both approximation algorithms considered
in the previous section cannot be improved.  Our examples are
unit-cost instances that are tight even for our modified greedy
algorithm and thus also for the greedy algorithm of Dolev et al.\
\cite{puzis+etal:prominent-group-complex-networks}.

\paragraph*{Tight Examples for \textsc{Maximum Coverage}}
Our examples are derived from worst-case examples of Khuller et al.\
\cite{Khuller+etal:budgeted-max-coverage} for unit-cost
\textsc{Maximum Coverage}.  These examples use a $(k+3)\times(k+1)$
matrix $(x_{ij})$ with $i=1,\ldots,k+3$ and $j=1,\ldots,k+1$ where $k$
is the number of sets to be selected.  For each row and for each
column there is a set in $\mathcal F$ that covers exactly the
respective matrix entries.  Only for column $j=k+1$ there is no
such set.

By a suitable choice of the weights $w(x_{ij})$ Khuller et al.\
achieve that in an optimal solution only rows are selected. On the
other hand, the greedy algorithm augments every initialization of
three sets (rows or columns) by choosing only columns during the
greedy steps.  (The example exploits that the greedy algorithm may
always choose columns in case of ties.)  They show that the output
produced this way has an approximation ratio arbitrarily close to
$1-1/e$ for high values of $k$.

\paragraph*{Tight Examples for MBC}
We simulate this construction by an instance of MBC.  We use that the
weights $w(x_{ij})$ of matrix entries can be written as
$w(x_{ij})=\alpha_{ij}/k^k$ where
\begin{displaymath}
\alpha_{ij}:=
\begin{cases}
	k^{k-j} (k-1)^{j-1} \quad & 1 \le j \le k\\
	(k-1)^{k} & j = k+1\,.
\end{cases}
\end{displaymath}
It should be clear that the example remains tight if we redefine
$w(x_{ij}):=\alpha_{ij}$ for any matrix entry $x_{ij}$.

\begin{figure}[ht]
	\begin{center}
		\includegraphics[width=\columnwidth]{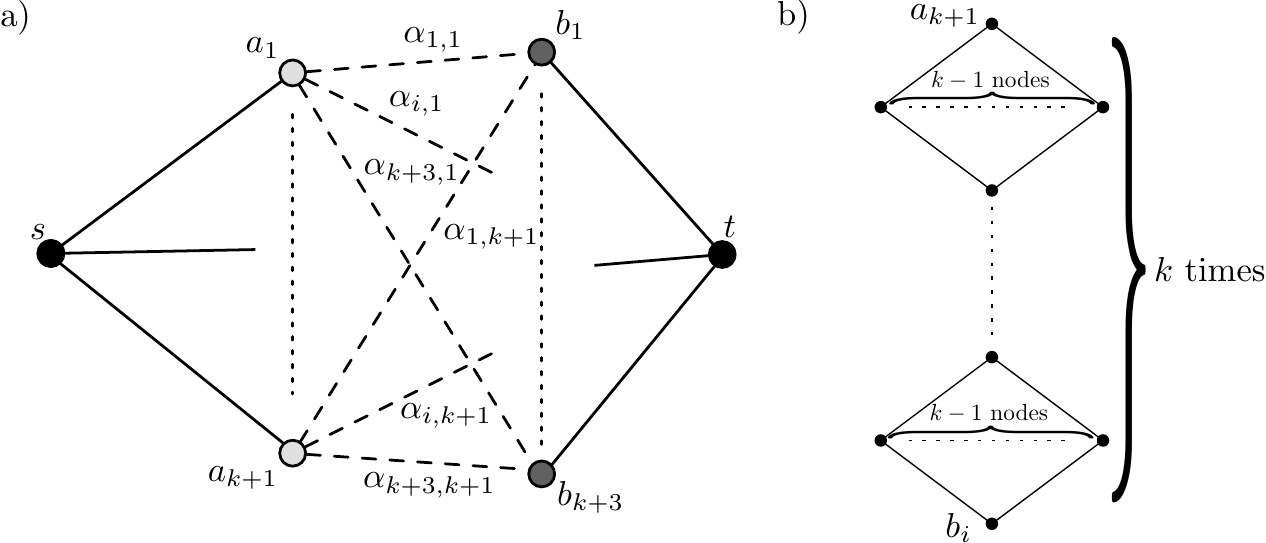}
	\end{center}
	\caption{a) construction of the tight examples; the dotted lines represent the
	nodes $a_{j}$ ($j= 1, \ldots, k+1$) and $b_{i}$ ($i = 1, \ldots, k+3$)
	respectively whereas the dashed lines mark the $a_{j}$--$b_{i}$ paths.
	b) con\-struc\-tion of the $a_{k+1}$--$b_{i}$ paths}
	\label{fig:konstruktionen}
\end{figure}

For our instance of MBC we introduce two distinguished nodes $s$ and $t$.  For
an illustration of our construction confer Figure~\ref{fig:konstruktionen}.
The basic idea is to represent every matrix entry $x_{ij}$ by exactly
$\alpha_{ij}$ shortest $s$--$t$ paths.  Each row $i$ is modeled by a node
$b_i$ and each column $j$ is modeled by a node $a_j$.  The set of shortest
$s$--$t$ paths meeting both $a_j$ and $b_i$ is exactly the set of shortest
$s$--$t$ paths representing $x_{ij}$.

For the sake of easier presentation we make some temporary
assumptions.  We explain later how those assumptions can be
removed. First we suppose only paths from vertex $s$ to vertex $t$
contribute to the group betweenness centrality.  Second, only the
vertices $a_{1}, \ldots, a_{k}$ and $b_{1}, \ldots, b_{k+3}$ are
candidates for the inclusion in a feasible solution $C\subseteq
V$. Note that the node $a_{k+1}$ should not be a candidate.

The $\alpha_{ij}$ shortest $a_j$--$b_i$ paths can be created by a
diamond like construction (Figure~\ref{fig:konstruktionen} shows this
construction for $\alpha_{i,k+1}$).

Recall that each node $b_i$ represents the row $i$ and each node $a_j$
represents column $j$.  Given our preliminary assumptions, it is clear
that in the above examples the feasible solutions for \textsc{Maximum
  Coverage} and MBC are in 1-1 correspondence.  Moreover,
corresponding solutions have the same goal function value. Hence the
modified greedy algorithm applied to the above instances produces
corresponding solutions for \textsc{Maximum Coverage} and MBC.  It
follows that the factor of $1-1/e$ is tight at least for the
restricted version of MBC that meets our preliminary assumptions.

\paragraph*{Removing the Preliminary Assumptions}
First, we drop the assumption that only $s$--$t$ paths are regarded.
We extend our schematic construction so that the shortest paths
between all pairs of vertices are considered but the matrix like
construction still works. We do this by replacing $s$ by a number
$l_s$ of vertices $s_{i}$ which are all directly linked with every
$a_j$ and all other $s_{i'}$. Similarly, $t$ is replaced by $l_t$
nodes $t_{j}$ directly linked with each $b_{i}$ and every other
$t_{j'}$. By increasing the numbers $l_{s}$ and $l_{t}$ we achieve
that only paths of the $s_{i}$--$t_{j}$ type are relevant.  This is
because the number of pairs $s_{i}, t_{j}$ is $\Omega(l_sl_t)$ whereas
the total number of remaining node pairs is $O(l_s+l_t)$.

Although we have achieved that only $s_i$--$t_j$ paths have
significant impact on the centrality of a solution $C$, we might face
problems if the numbers of covered $s_i$--$t_j$ paths are equal for two
feasible solutions.  This is because we have assumed that the greedy
algorithm chooses columns (or nodes $a_j$) in case of ties regarding
only $s_i$--$t_j$ paths.  We can resolve this issue by making $l_{s}$ greater
than $l_{t}$; this ensures that during the greedy steps always one of
the $a_{i}$ nodes is preferred.

The remaining problem is to ensure that only the nodes $a_{1}, \dots,
a_{k}$ and $b_{1}, \dots, b_{k+3}$ are allowed to be part of a
solution.  First we exclude $a_{k+1}$ as a candidate.  This is
accomplished by splitting $a_{k+1}$ into multiple nodes, so that every
$b_{i}$ has its own node $a_{k+1,i}$. The node $a_{k+1,i}$ is linked
by an edge with each $s_{i}$ and by $\alpha_{k+1,i}$ paths with the
node $b_{i}$. As all $s_{i'}$--$t_{j'}$ paths covered by $a_{k+1,i}$
are also covered by $b_{i}$ we may assume that none of the nodes
$a_{k+1,i}$ is used by a solution.  Now consider a node $u$ that lies
on some shortest $a_j$--$b_i$ path.  It can be observed that $a_j$
covers any shortest $s_{i'}$--$t_{j'}$ path that is covered by $u$.
Therefore we may prefer $a_j$ over $u$.  Finally consider a node
$s_i$.  Then the centrality of $s_i$ is $O(l_s+l_t)$ whereas the
centrality of any node $a_j$ is $\Omega(l_sl_t)$.  It follows that
only the nodes $a_{1}, \dots, a_{k}$ and $b_{1}, \dots, b_{k+3}$ are
relevant candidates for the inclusion in a good solution.

As all preliminary assumptions can be removed, we get

\begin{theorem}
  The approximation factor of $1-1/e$ of the greedy algorithm for MBC
  is tight.
	\label{theorem:approx_kpp_tight}
	\qed{}
\end{theorem}

Our construction uses unit weights only. As the
modified greedy algorithm starts the greedy procedure for \emph{every}
subset $C \subseteq V$ of at most three vertices, its output cannot be
worse than the output of the simpler greedy algorithm of Dolev et al.\
\cite{puzis+etal:prominent-group-complex-networks}. Hence the
approximation factor of $1-1/e$ of their algorithm is also tight.

\section{APX-completeness}\label{sec:apx-completeness}

In this section we prove that unit-cost MBC is APX-complete
thereby showing that it does not admit a PTAS on general graphs.

We do this by giving an approximation preserving reduction from
\textsc{Maximum Vertex Cover}.  This problem is defined as follows. We
are given an undirected graph $G=(V,E)$ and a number $k$.  We are
looking for a $k$-element node set $V'$ such that the number of edges
that are incident at some node in $V'$ is maximum.  \textsc{Maximum
  Vertex Cover} is known to be APX-complete
\cite{DBLP:journals/cc/Patrank94}.

Our proof consists of several steps.  First we describe a polynomial
time transformation of an instance $(G,k)$ of \textsc{Maximum Vertex
  Cover} to an instance $(G',k)$ of MBC.  Then we introduce a
modified centrality measure $\GBC'$ for which it is easier to
establish a correspondence between (approximate) solutions of MBC and
\textsc{Maximum Vertex Cover}.  We argue that it is sufficient to
consider this modified measure instead of the betweenness centrality.
Finally, we observe that for any (relevant) node set $C$ its modified
centrality $\GBC'$ in $G'$ and the number of edges covered by $C$ in
$G$ are proportional which completes the proof.

\paragraph*{The Transformation} Given an instance $(G,k)$ of
\textsc{Maximum Vertex Cover} we construct a graph $G'$ that contains
all nodes of $V$ and additionally for each $v\in V$ a set
$v_1,\ldots,v_l$ of \emph{copies} of $v$.  Here, $l$ is a large number
to be chosen later.

Now we specify the edge set of $G'$.  First we connect for each $v\in
V$ the node set $\{v,v_1,\ldots,v_l\}$ to a clique with $l+1$ nodes.
Let $u,v$ be two distinct nodes in $V$.  If $u$ and $v$ are adjacent
in $G$ then they are so in $G'$.  If $u$ and $v$ are \emph{not}
adjacent in $G$ then we introduce an intermediate node $z_{uv}$ and
connect each $u_i$ and each $v_j$ with $z_{uv}$ where
$i,j=1,\ldots,l$.  The number $k$ represents the cardinality of the
solution in both instances.  This completes the construction of $G'$.

\paragraph*{Modified Centrality} Any pair $(u_i,v_j)$ of copies of
\emph{distinct} nodes $u,v\in V$ is called \emph{essential}.  The
remaining node pairs in $G'$ are \emph{inessential}.

We are able to show that it suffices to work with the \emph{modified group
betweenness centrality}
\[\gbc'(C) := \sum_{(u_i,v_j)\text{ is essential}}
\frac{\sigma_{u_i,v_j}(C)}{\sigma_{u_i,v_j}}\]
 that is, to respect only essential
node pairs.  The basic reason for this is that for any node set $C$
the total contribution of inessential node pairs to the centrality
measure $\GBC$ is linear in $l$.  On the other hand, the contribution
of essential pairs to reasonable solutions is always at least $l^2$
since the inclusion of at least one node $u\in V$ into $C$ already
covers all $l^2$ shortest $u_i$--$v_j$ paths for any $v$ adjacent to
$u$ in $G$.  Therefore we can make the impact of inessential pairs
arbitrarily small by choosing $l$ large enough.

\paragraph*{Reduction from \textsc{Maximum Vertex Cover} to MBC} Now
we show that our above transformation of $G$ to $G'$ can in fact be
extended to an approximation preserving reduction from \textsc{Maximum
  Vertex Cover} to the modified centrality problem.  That is we have
to specify how a feasible solution for the latter problem can be
transformed back into a solution for \textsc{Maximum Vertex Cover}
that preserves the approximation ratio.

To this end consider an arbitrary node set $C$ of $V'$.  If $C$
already covers all edges in $G$ we are finished.  Otherwise there is
an edge $(u,v)$ that is not covered by $C$.  Now assume that $C$
contains a copy $u_i'$ of some node $u'\in V$.  The only essential
shortest paths that are occupied by $u_i'$ are $O(nl)$ shortest
paths to copies $v_j'$ of nodes $v'\in V$ that are not adjacent to
$u'$ in $G$.  Now suppose that we replace node $u_i'$ in $C$ with node
$u$ of the uncovered edge $(u,v)$.  Then $u$ covers at least $l^2$
previously uncovered shortest $u_i$--$v_j$ paths between copies of $u$
and $v$, respectively.  Thus if $l$ was chosen to be large in
comparison to $n$ the modified centrality can only increase under
this replacement.

If $C$ contains an intermediate node $z_{u'v'}$ then this node covers
exactly $l^2$ shortest $u_i'$--$v_j'$ paths.  Hence the modified
centrality does not decrease if we replace $z_{u'v'}$ with $u$.

To summarize we have shown how we can transform any node set $C$ in
$G'$ into a node set for $G$ without decreasing the modified
centrality.  In other words we can restrict our view to node subsets
of $V$.  Now consider such a node set $C$ that contains only nodes of
$V$.  It is easy to verify that $C$ covers exactly all shortest
$u_i$--$v_j$ paths of edges $(u,v)$ in $G$ for which at least one end
point lies in $C$.  In other words the modified centrality of $C$
equals the number of edges covered by $C$ multiplied with exactly
$l^2$.  Hence the measures for \textsc{Maximum Vertex Cover} and the
modified MBC are proportional. This completes the
reduction from \textsc{Maximum Vertex Cover} to the modified
centrality problem.

\begin{theorem}
  Unit-cost MBC is APX-complete. \qed{}
\end{theorem}

\section{A Polynomial-Time Algorithm for Trees}\label{sec:polyn-time-algor}

We complement the hardness result for general graphs of the previous
section by a tractable special case.  Specifically, we show that the
budgeted MBC problem can be solved efficiently on trees using a
dynamic programming approach.

Let $T=(V,E)$ be a tree.  We assume that $T$ is rooted at some
arbitrary node $r$.  If $v$ is a node in $T$ then $T_v$ denotes the
subtree of $T$ hanging from~$v$.

Let $s,t$ be an arbitrary pair of distinct nodes of the tree $T$.
Since $T$ contains exactly one $s$--$t$ path, we have $\sigma_{s,t}=1$.
Let $C\subseteq V$ be a set of nodes. Then $\sigma_{s,t}(C)=1$ if the
$s$--$t$ path contains some node from $C$, and otherwise
$\sigma_{s,t}(C)=0$.  Thus the betweenness centrality $\gbc (C)$ of
$C$ simplifies greatly.  It equals the number of $s$--$t$ pairs ($s$
and $t$ always distinct) \emph{covered} by $C$ (meaning
$\sigma_{s,t}(C)=1)$.

Our dynamic program uses a three-dimensional table $B$ whose entries we now
define. Let $v$ be some node in $T$, let $\sigma \leq n^2$ be a non-negative
integer value, and let $m\leq |T_v|$.  Then $B[v,\sigma,m]$ denotes the cost of
the cheapest node set $C\subseteq T_v$ with the following two properties.
\begin{enumerate}[(i)]
\item $\gbc_v(C)\geq \sigma$ where $\gbc_v(C)$ denotes the number of
  $s$--$t$ pairs in $T_v$ covered by $C$.
\item There are at least $m$ nodes $u$ (including $v$) in $T_v$ such
  that the $u$--$v$ path is not covered by $C$.  We call such nodes
  \emph{top nodes} of $T_v$.
\end{enumerate}

In what follows we describe how those $B[\cdot]$-values can be
computed in polynomial time in a bottom-up fashion.  The optimum value
of $\GBC$ in the input tree $T$ then equals the maximum value
$\sigma\leq n^2$ such that $B[r,\sigma,0]\leq b$.  We explain our
algorithm for \emph{binary} trees.  The general case can essentially
be reduced to the case of binary trees by splitting any node with
$k\geq 3$ children into $k-1$ binary nodes.

Consider a node $v$ with children $v_1$ and $v_2$.  We wish to compute
$B[v_,\sigma,m]$.  Assume by inductive hypothesis that we already know
all values $B[v_i,\cdot,\cdot]$ for $i=1,2$.

Suppose first that $m\geq 1$, which implies $v\notin C$.  Let $m_i$ be
the number of top nodes in $T_{v_i}$. Then $m_1+m_2+1\geq m$.
Altogether there are $\bar{\sigma}:=(|T_{v_1}|+1)(|T_{v_2}|+1)-1$ many
$s$--$t$ pairs such that $s$ and $t$ do not lie in the same subtree
$T_{v_i}$. It is exactly those pairs of $T_v$ nodes that have not yet
been accounted for within the subtrees $T_{v_i}$.  Such a pair is
\emph{not} covered if and only if $s$ and $t$ are both top nodes of
$T_v$.  There are $(m_1+1)(m_2+1)-1$ such pairs.  Hence the number of
covered node pairs $s$, $t$ such that $s$ and $t$ do not lie in the
same subtree $T_{v_i}$ is given by $\bar{\sigma}(m_1,m_2):=
\bar{\sigma}-(m_1+1)(m_2+1)-1$. The value $B[v,\sigma,m]$ is given by the
minimum of the values $B[v_1,\sigma_1,m_1]+B[v_2,\sigma_2,m-m_1-1]$
such that $\sigma_1+\sigma_2+\bar{\sigma}(m_1,m-m_1) = \sigma$.
Therefore $B[v,\sigma,m]$ can be computed in $O(m\sigma)=O(n^3)$ time.

Now consider the case $m=0$.  If $v\notin C$ then we can proceed as in
the case $m=1$.  If $v\in C$ then any of the $\bar{\sigma}$ pairs
$s,t$ with $s$ and $t$ not in the same subtree is covered by $C$.
Hence, if $v\in C$, then $B[v,0,\sigma]$ equals the minimum $\bar{B}$
of the values $c(v)+B[v_1,\sigma_1,0]+B[v_2,\sigma_2,0]$ such that
$\sigma_1+\sigma_2+\bar{\sigma} = \sigma$, which can be computed in
$O(\sigma)=O(n^2)$ time.  Altogether we have that
$B[v,\sigma,0]=\min\{\bar{B},B[v,\sigma,1]\}$.

Finally, if $v$ is a leaf then $B[v,0,m]=0$ for $m=0,1$.

Since there are $O(n^4)$ values $B[v,\sigma,m]$ each of which can be
computed in $O(n^3)$ we obtain a total running time of $O(n^7)$ for
computing the optimum budgeted betweenness centrality on a binary
tree.

\begin{theorem}\label{thm:polyn-time-algor-trees}
  The budgeted MBC problem can be solved in polynomial time on a
	tree. \qed{}
\end{theorem}

\section{Concluding Remarks}

We have introduced a reduction from MBC to \textsc{Maximum Coverage}
that allows us to simplify the analysis of the greedy approach of
Dolev et al.\
\cite{dolev+etal:incremental-deployment-group-betweenness} for the
unit-cost version and to derive a new algorithm for a budgeted
generalization of MBC.  We have provided a class of tight examples for
both algorithms.  Moreover, we have shown that MBC is APX-complete but
can be solved in polynomial time on trees.

Our reduction suggests to consider MBC as a special case of
\textsc{Maximum Coverage}.  It is well-known that \textsc{Maximum
  Coverage} cannot be approximated strictly better than $1-1/e$ unless
$\text{P}=\text{NP}$ \cite{feige:set-cover-threshold}.  However, it
seems to be difficult to derive a similar upper bound for MBC since
the \textsc{Maximum Coverage} instances corresponding to MBC have a
very specific structure.  As there is at least one shortest path for
any pair of nodes in a connected graph, the number $|\mathcal F|$ of
sets in the \textsc{Maximum Coverage} instance is $O(\sqrt{|S|})$
where $S$ is the set of ground elements.  

On the other hand, the best known algorithm for \textsc{Maximum Vertex
  Cover}, developed by Ageev and Sviridenko
\cite{ageev+srinivasan:IPCO99:max-coverage}, has a ratio of $3/4$.
Our approximation preserving reduction from \textsc{Maximum Vertex
  Cover} to MBC provided in Section~\ref{sec:apx-completeness} shows
that a significantly better approximability result for MBC would also
imply a better approximation for \textsc{Maximum Vertex Cover}.
Conversely, this reduction suggests to try the techniques of Ageev and
Sviridenko \cite{ageev+srinivasan:IPCO99:max-coverage} as possible
avenues to improve the approximation factor for MBC.

\newpage
\appendix
{\huge{\textbf{\appendixname}}}

\section{Justification of the Modified Betweenness Centrality}

Recall that we used in the proof of the APX-completeness in
Section~\ref{sec:apx-completeness} the modified centrality
\[\gbc'(C) := \sum_{(u_i,v_j)\text{ is essential}}
\frac{\sigma_{u_i,v_j}(C)}{\sigma_{u_i,v_j}}\]
 instead of $\GBC$.  In order to
justify this more formally, we give an approximation preserving
reduction from the modified problem version to MBC.

Let $\OPT$ and $\OPT'$ denote the optimum centrality for the problem
instance $G'$ (for the construction of $G'$ confer
Section~\ref{sec:apx-completeness}) with respect to $\GBC$ and
$\GBC'$, respectively.  Consider a $k$-element node set $C$ such that
$\GBC(C)\geq (1-\epsilon)\OPT$.  We claim that $\GBC'(C)\geq
(1-2\epsilon)\OPT'$ if $l$ was chosen large enough.  This completes
the reduction from the modified problem version to the original one.

The claim can be seen as follows: The first type of inessential node
pairs form pairs $(v_i,v_j)$ of copies of the same node $v\in V$.  The
only shortest path between $v_i$ and $v_j$ is the direct connection.
Hence any node in $C$ occupies at most $l-1$ of such paths.  This
implies that the centrality of $C$ drops by at most $O(nl)$ when we
ignore inessential node pairs of the first type.

The second type of inessential node pairs form pairs $(z,z')$ where at
least one of the nodes $z$ and $z'$ is not a copy of a node in $G$.
In other words, this node is either a node in $G$ or an intermediate
node $z_{uv}$ for some edge $(u,v)$ in $G$.  Since there are only
$O(m^2l)$ inessential pairs of this type the absolute error we
make when switching to the modified betweenness centrality is bounded
by $c m^2l$ for some constant $c$, that is,
$\GBC'(C)\geq\GBC(C)-cm^2l$.

Let $(u,v)$ be some edge in $G$.  We can cover at least all $l^2$
shortest $u_i$--$v_j$ paths in $G'$ by including $u$ into our solution
$C$.  This implies $\OPT'\geq l^2$.  By choosing $l\geq
(cm^2)/\epsilon$ we can ensure that our solution $C$ has a modified
centrality $\GBC'(C)$ of at least $(1-\epsilon)\OPT-cm^2l\geq
(1-2\epsilon)\OPT'$ as desired.

\section{Polynomial Time Algorithm for Trees of Arbitrary Degree}

In Section~\ref{sec:polyn-time-algor} we have provided a polynomial
time algorithm for solving MBC on \emph{binary} trees.

As we remarked the case of arbitrary trees can essentially be reduced
to the case of a binary tree.  To this end consider a node $v$ with
children $v_1,\ldots,v_k$.

The case $k=1$ can be handled similarly to $k=2$ and is in fact
easier.  If $m\geq 1$ then $B[v,\sigma,m]$ equals $B[v_1,\sigma,m-1]$. If $m=0$
then $B[v,\sigma,0]$ is the minimum of $B[v,\sigma,1]$ and
$c(v)+B[v_1,\sigma-|T_{v_1}|,0]$.

If $k\geq 3$ we face the problem that there are possibly exponentially
many ways of distributing the $m$ top nodes to the subtrees $T_{v_i}$.
To overcome this difficulty we split $v$ into $k-1$ binary nodes.
More precisely, we introduce a set $U(v)$ of $k-1$ new nodes
$u_1,\ldots,u_{k-1}$ and replace $v$ and the edges incident at $v$
with the edge set $\{\,(u_i,v_i),(u_i,u_{i+1})\mid
i=1,\ldots,k-1\,\}$.  Here we set $u_k=v_k$.  The cost $c(u_{k-1})$
is set to $c(v)$ the remaining costs $c(u_i)$ are zero.

Now we can treat these newly introduced nodes very similarly to the
binary nodes of the original tree.  The difference is that we need to
handle the nodes in $U(v)$ as a single top node and as a single end
node of paths.  Moreover, we have to ensure that either all of the
nodes in $U(v)$ are included in $C$ or none of them.  (One can picture
the $u_1$--$u_{k-1}$ path as an expanded version of the originally
single node $v$.)  

To this end we handle $u_{k-1}$ like a regular binary node as
described above.  Now consider $u_i$ with $i\leq k-2$ having children
$v_i$ and $u_{i+1}$.  If $m\geq 1$ and hence $u_i\notin C$ then
$B[u_i,\sigma,m]$ equals the minimum value
$B[v_i,m_1,\sigma_1]+B[u_{i+1},m_2,\sigma_2]$ such that $m_1+m_2=m$,
$m_2\geq 1$ and
$\sigma_1+\sigma_2+|T_{v_i}|\cdot( |T_{u_{i+1}}-U(v)| + 1) - m_1 m_2 =
\sigma$.  We require that $m_2\geq 1$ since we have to ensure that
either all of the nodes in $U(v)$ are included in $C$ or none of them.

Now consider the case $m=0$.  For $i = 1, \ldots, k-2$ let $\bar{B}_{i}$ be the
minimum value $B[v_i,\sigma_1,0]+B[u_{i+1},\sigma_2,0]$ such that
$\sigma_1+\sigma_2+|T_{v_i}|(|T_{u_{i+1}}-U(v)|+1) = \sigma$. We have to
ensure that only $B[\cdot]$-values are combined in which the inclusion of $v$ in
a central node set $C$ (i.e.\ $m=0$) is assumed either for all $u_{i}$ ($i = 1,
\ldots, k-1$) or for none. Therefore, the only node for which we include the
case $m \ge 1$ in the case $m=0$ is $u_{1}$ (remember that $m$ is only a lower
bound for the number of top nodes). Thus $B[u_{1}, \sigma, 0]$ equals $\min
\left\{ \bar{B}_{1}, B[u_{1}, \sigma, 1] \right\}$. For $2 \le i \le k-2$ we get
$B[u_i,\sigma,0] = \bar{B}_{i}$. We also have to ensure that for $u_{k-1}$
the cost $B[u_{k-1}, \sigma, 1]$ is not considered during the computation of
$B[u_{k-1}, \sigma, 0]$ which leads to $B[u_{k-1}, \sigma, 0] = \bar{B}$ where,
as in Section \ref{sec:polyn-time-algor}, $\bar{B}$ equals the minimum of the
values $c(v)+B[v_{k-1},\sigma_1,0]+B[v_{k},\sigma_2,0]$ such that
$\sigma_1+\sigma_2 + (|T_{v_{k-1}}| + 1)(|T_{v_{k}}| + 1) - 1 = \sigma$.  All of
the above computations can be carried out in $O(n^3)$ per value
$B[u_i,\sigma,m]$.

Finally, we observe that the number of nodes can at most double by the
above splitting construction.  Which yields
Theorem~\ref{thm:polyn-time-algor-trees}.


\begin{thebibliography}{10}

\bibitem{ageev+srinivasan:IPCO99:max-coverage}
A.~A. Ageev and M.~I. Sviridenko.
\newblock Approximation algorithms for maximum coverage and max cut with given
  sizes of parts.
\newblock In {\em Proceedings of 7th Conference on Integer Programming and
  Combinatorial Optimization (IPCO'99)}, volume 1610 of {\em Lecture Notes in
  Computer Science}, pages 17--30, 1999.

\bibitem{brandes:faster-algorithm-centrality}
U.~Brandes.
\newblock {A faster algorithm for Betweenness Centrality}.
\newblock {\em Journal of Mathematical Sociology}, 25(2):163--177, 2001.

\bibitem{brandes:variants-of-betweenness}
U.~Brandes.
\newblock {On variants of Shortest-Path Betweenness Centrality and their
  generic computation}.
\newblock {\em Social Networks}, 30(2):136--145, 2008.

\bibitem{dolev+etal:incremental-deployment-group-betweenness}
S.~Dolev, Y.~Elovici, R.~Puzis, and P.~Zilberman.
\newblock Incremental deployment of network monitors based on group betweenness
  centrality.
\newblock {\em Information Processing Letters}, 109(20):1172--1176, 2009.

\bibitem{everett-borgatti:centrality-groups}
M.~Everett and S.~Borgatti.
\newblock {The centrality of groups and classes}.
\newblock {\em Journal of Mathematical Sociology}, 23:181--202, 1999.

\bibitem{feige:set-cover-threshold}
U.~Feige.
\newblock A threshold of $\ln n$ for approximating set cover.
\newblock {\em Journal of the ACM}, 45(4):634--652, 1998.

\bibitem{freeman:freeman-measures-centrality}
L.~Freeman.
\newblock {A set of measures of centrality based on betweenness}.
\newblock {\em Sociometry}, 40(1):35--41, 1977.

\bibitem{Khuller+etal:budgeted-max-coverage}
S.~Khuller, A.~Moss, and J.~Naor.
\newblock The budgeted maximum coverage problem.
\newblock {\em Information Processing Letters}, 70:39--45, 1999.

\bibitem{newman:shorest-path-centrality}
M.~E.~J. Newman.
\newblock {Scientific collaboration networks. II. Shortest paths, weighted
  networks, and centrality}.
\newblock {\em Physical Review E}, 64:016132, 2001.

\bibitem{DBLP:journals/cc/Patrank94}
E.~Petrank.
\newblock The hardness of approximation: Gap location.
\newblock {\em Computational Complexity}, 4:133--157, 1994.

\bibitem{puzis+etal:successive-computation-group-betweenness-centrality}
R.~Puzis, Y.~Elovici, and S.~Dolev.
\newblock Fast algorithm for successive computation of group betweenness
  centrality.
\newblock {\em Phys. Rev. E}, 76(5):056709, Nov 2007.

\bibitem{puzis+etal:prominent-group-complex-networks}
R.~Puzis, Y.~Elovici, and S.~Dolev.
\newblock {Finding the most prominent group in complex networks}.
\newblock {\em AI Communications}, 20(4):287--296, 2007.

\end{thebibliography}
\end{document}